\documentclass[aps,prl,twocolumn,floatfix]{revtex4}
\usepackage{graphicx,amsmath,xspace,CJK}

\begin{document}
\begin{CJK*}{GB}{}
\title{Glass transitions in two-dimensional suspensions of colloidal ellipsoids}

\author{Zhongyu Zheng, Feng Wang and Yilong Han$^{*}$}
\affiliation{Department of Physics, Hong Kong University of Science and
 Technology, Clear Water Bay, Hong Kong, China}
\email{yilong@ust.hk}
\date{\today}

\begin{abstract}
We observed a two-step glass transition in monolayers of colloidal ellipsoids by video microscopy. The glass transition in the rotational degree of freedom was at a lower density than that in the translational degree of freedom. Between the two transitions, ellipsoids formed an orientational glass. Approaching the respective glass transitions, the rotational and translational fastest-moving particles in the supercooled liquid moved cooperatively and formed clusters with power-law size distributions. The mean cluster sizes diverge in power law as approaching the glass transitions. The clusters of translational and rotational fastest-moving ellipsoids formed mainly within pseudo-nematic domains, and around the domain boundaries, respectively.
\end{abstract}

\maketitle
\end{CJK*}

Colloids are outstanding model systems for glass transition studies because the trajectories of individual particles are measurable by video microscopy \cite{Weeks00}. In the past two decades, significant experimental effort has been applied to studying colloidal glasses consisting of isotropic particles \cite{Gotze91,Megen93,Weeks00,Kegel00,Zhang09}, but little to anisotropic particles \cite{Yunker11}. The glass transition of anisotropic particles has been studied in three dimensions (3D) mainly through simulation \cite{Stillinger94,Yatsenko08}. Molecular mode-coupling theory (MMCT) predicts that particle anisotropy should lead to new phenomena in glass transitions \cite{Letz00,Schilling97}, and some of these have been observed in recent 3D simulations of hard ellipsoids \cite{Michele07,Pfleiderer08}.
MMCT \cite{Franosch97,Schilling97} suggests that hard ellipsoids with an aspect ratio $p>2.5$ in 3D can form an orientational glass in which rotational degrees of freedom become glass while the center-of-mass motion remains ergodic \cite{Letz00}. Such a ``liquid glass"~\cite{Schilling00}, in analogy to a liquid crystal, has not yet been explored in 3D or even 2D experiments. Anisotropic particles should also enable exploration of the dynamic heterogeneity in the rotational degrees of freedom. Moreover the glass transitions of monodispersed particles have not yet been studied in 2D. It is well known that monodispersed spheres can be quenched to a glass in 3D, but hardly in 2D even at the fastest accessible quenching rate. Hence bidispersed or highly polydispersed spheres have been used in experiments \cite{Konig05, Zhang09,Yunker09}, simulations \cite{Speedy99} and theory \cite{Bayer07} for  2D glasses. In contrast, we found that monodispersed ellipsoids of intermediate aspect ratio are excellent glass formers in 2D because their shape can effectively frustrate crystallization and nematic order.

Here we investigate the glass transition in monolayers of colloidal ellipsoids using video microscopy. We measured the translational and rotational relaxation times, the non-Gaussian parameter of the distribution of displacements, and the clusters of cooperative fastest-moving particles. These results consistently showed that the glass transitions of rotational and translational motions occur in two different area fractions, defining an intermediate orientational glass phase.

The ellipsoids were synthesized by stretching polymethyl methacrylate (PMMA) spheres~\cite{Ho93,Han09}. They had a small polydispersity of 5.6\% with the semi-long axis $a = 3.33~\mu$m and the semi-short axes $b = c = 0.56~\mu$m. 3mM sodium dodecyl sulfate (SDS) was added to stabilize ellipsoids and the $>3$~mM ionic strength in the aqueous suspension made ellipsoids moderately hard particles. A monolayer of ellipsoids was strongly confined between two glass walls \cite{Han09}. Light interference measurements showed that the wall separation varied by only $\sim$30~nm per 1~mm \cite{Han09}, so the walls could be considered as parallel within the field of view. The area fraction $\phi \equiv \pi ab\rho$, where $\rho$ is the number density averaged over all video frames. Twelve densities were measured in the range $0.20\le \phi \le 0.81$. During the three to six hours measurments at each $\phi$, no drift flow or density change was observed. The thermal motion of the ellipsoids was recorded using a charge-coupled device camera resolving 1392$\times$1040 pixels at 1 frame per second (fps) for the highest five concentrations and at 3 fps for lower concentrations. The center-of-mass positions and orientations of individual ellipsoids were tracked using our image processing algorithm \cite{Zheng10}. The angular resolution was 1$^\circ$ and the spatial resolutions were 0.12 $\mu$m and 0.04 $\mu$m along the long and the short axes respectively. More experimental details are in the Supplemental Material (SM).

\begin{figure}
\begin{center}
\includegraphics[width=1\columnwidth]{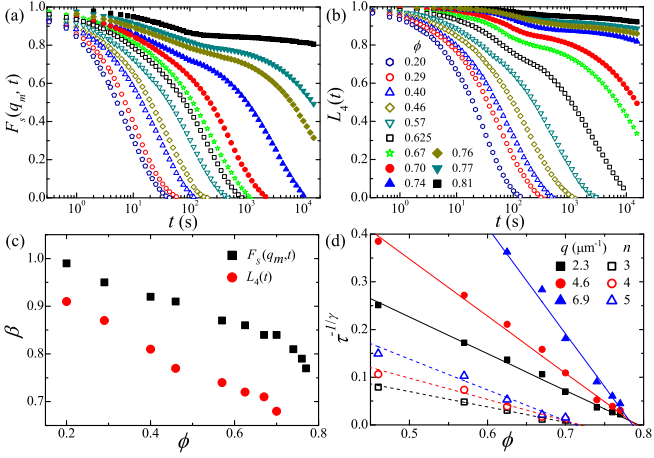}
\caption{(color online) (a) The self-intermediate scattering function $F_s(q,t)$ at $q_m=2.3~\mu$m$^{-1}$ and (b) the orientational correlation $L_4(t)$ for different area fractions. (c) The exponent $\beta$ of the fitting function $e^{-(t/\tau)^{\beta}}$ for the long-time $F_s(q_m,t)$ and $L_4(t)$. (d) The fitted relaxation time $\tau(\phi) \sim (\phi_c-\phi)^{-\gamma}$. Solid symbols: different choices of $q$ in $F(q,t)$ for the translational motion. Open symbols: different choices of $n$ in $L_n(t)$ for the orientational motion. } \label{fig:Fq}
\end{center}
\end{figure}
At high densities the ellipsoids spontaneously formed small pseudo-nematic domains with branch-like structures each involving about $10^2$ particles, see Fig.~S1 of the SM. The translational relaxation was characterized by the self-intermediate scattering function $F_s(q,t)\equiv \langle \sum_{j=1}^N e^{i\mathbf{q}\cdot (\mathbf{x}_j(t) -\mathbf{x}_j(0))} \rangle/N$ where $\mathbf{x}_j(t)$ is the position of ellipsoid $j$ at time $t$, $N$ is the total number of particles, $\mathbf{q}$ is the scattering vector and $\langle ~\rangle$ denotes a time average. In Fig.~\ref{fig:Fq}(a), we chose $q_m=2.3~\mu$m$^{-1}$ measured from the first peak position in the structure factor at high density. The rotational relaxation can be characterized by the $n^{\textrm{th}}$ order of the orientational correlation function $L_n(t) \equiv \langle \sum_{j=1}^N\cos{n(\theta_j(t)-\theta_j(0))} \rangle/N$ where $n$ is a positive integer and $\theta_j$ is the orientation of ellipsoid $j$. $L_n(t)$ decays faster for larger $n$, and different choices of $n$ can yield the same glass transition point. $n=4$ in Fig.~\ref{fig:Fq}(b) was chosen so that $L_n(t)$ can be better displayed within our measured time scales. At high $\phi$, both $F_s(q_m,t)$ and $L_4(t)$ develop two-step relaxations, the characteristics upon approaching the glass transition. The short-time $\beta$-relaxation corresponds to motion within cages of neighboring particles, and the long-time $\alpha$-relaxation reflects structural rearrangement involving a series of cage breakings. According to mode-coupling theory (MCT), the $\alpha$-relaxation follows $e^{-(t/\tau)^{\beta}}$. Figure~\ref{fig:Fq}(c) shows that $\beta$ decreases with density, indicating dynamic slowing down upon supercooling \cite{Kawasaki07,Michele07}.

MCT predicts that the relaxation time $\tau(\phi)$ diverges algebraically approaching the critical point $\phi_c$: $\tau(\phi) \sim (\phi_c-\phi)^{-\gamma}$ where $\gamma=1/(2a)+1/(2b)$ \cite{Gotze92}. Here $a$ and $b$ are the exponents in the critical-decay law $F_s(q,t) =f_q^{c}+h_qt^{-a}$ and the von Schweidler law $F_s(q,t) =f_q^{c}-h_qt^{b} $ at the initial stage of the $\beta$-relaxation and the crossover time to the $\alpha$-relaxation respectively. The fitted $a$ or $b$ is almost a constant at different $\phi$, indicating that $F_s(q,t)$ can collapse onto a master curve in the appropriate time regime. This demonstrates that $F_s(q, t)$ can be separated into a $q$-dependent and a $t$-dependent part \cite{Gotze92}. Interestingly, $L_n(t)$ can similarly collapse. The fitted $a_T=0.3\pm0.02$  and $b_T=0.63\pm0.02$ for $F_s(q_m, t)$ and  $a_{\theta}=0.32\pm0.02$  and $b_{\theta}=0.65\pm0.02$ for $L_4(t)$ yield $\gamma_T=2.45\pm0.05$ and $\gamma_{\theta}=2.33\pm0.05$ for the translational and orientational correlations respectively. These values are close to the $\gamma_T=2.3$ measured for 3D ellipsoids \cite{Pfleiderer08}. In Fig.~\ref{fig:Fq}(d), $\tau^{-1/\gamma}$ is linear in $\phi$ for different choices of $q$ and $n$. Interestingly, all the scalings show that the glass transitions are at  $\phi_c^{\theta}=0.72\pm0.01$ for rotational motion and $\phi_c^{_T}=0.79\pm0.01$ for translational motion.  This indicates three distinct phases: liquid ($\phi<0.72$), an intermediate orientational glass which is liquid-like in its translational degrees of freedom but glassy in its rotational degrees of freedom ($0.72<\phi<0.79$), and the glass state for both degrees of freedom ($\phi>0.79$).

\begin{figure}
\begin{center}
\includegraphics[width=1\columnwidth]{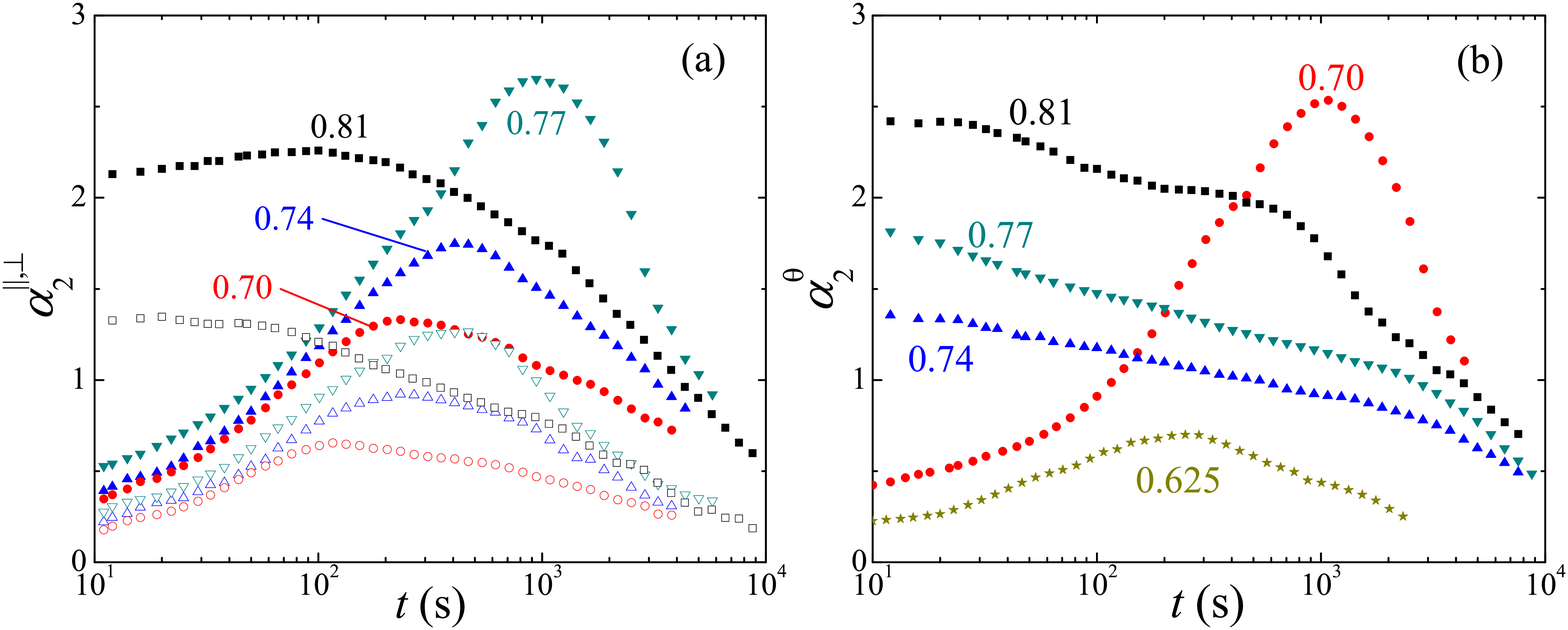}
\caption{ (color online) (a) The non-Gaussian parameters of translational displacements along the long axis ($\alpha_2^{||}(t)$, solid symbols) and the short axis ($\alpha_2^{\perp}(t)$, open symbols). $\phi = 0.70, 0.74, 0.77, 0.81$ as labeled in the figures. (b) The non-Gaussian parameters of rotational displacements.} \label{fig:nonGauss}
\end{center}
\end{figure}
Besides the extrapolations in Fig.~\ref{fig:Fq}(d), the existence of the orientational glass phase was verified from the non-Gaussian parameters $\alpha_2(t)=\langle \Delta x^4 \rangle/(3\langle \Delta x^2 \rangle^2)-1$ of particle displacements $\Delta x$ during time $t$ \cite{Weeks00}. In supercooled liquids, the distribution of $\Delta x$ is Gaussian at short and long times because the motions are diffusive, but it becomes non-Gaussian with long tails at the intermediate times due to cooperative out-of-cage displacements \cite{Weeks00,Kegel00,Donati99}. This behavior is reflected in the peak of  $\alpha_2(t)$, see Fig.~\ref{fig:nonGauss}. As $\phi$ increases, the peak rises and shifts towards a longer time, indicating growing dynamic heterogeneity on approaching the glass transitions. In contrast, the glass phase lacks cooperative out-of-cage motions, so $\alpha_2(t)$ exhibits no distinct peak and declines with time~\cite{Weeks00}. Such a sharp change has been regarded as a characteristic of a glass transition \cite{Weeks00}. Figure~\ref{fig:nonGauss} clearly shows the glass transitions at $\phi_c^{\theta}=0.72\pm 0.02$ for rotational motion and at $\phi_c^{_T}=0.79\pm 0.02$ for translational motion. In Fig.~\ref{fig:nonGauss}(a), $\alpha_2^{||}(t)$ is always greater than the corresponding $\alpha_2^{\perp}(t)$, indicating that the translational relaxations and cooperative out-of-cage motions are mainly along the long axes of the ellipsoids.

\begin{figure}
\begin{center}
\includegraphics[width=1\columnwidth]{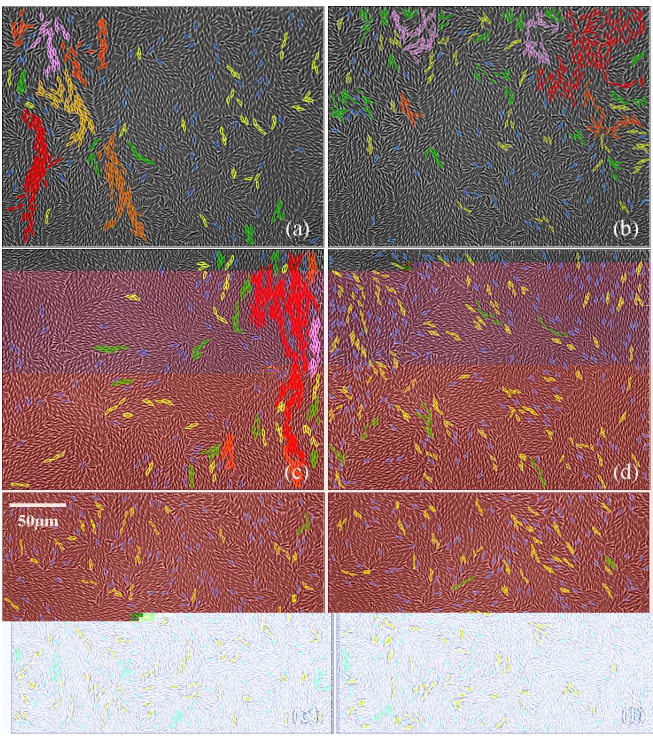}
\caption{(color online) The spatial distributions of the fastest-moving 8\% of the particles (labeled in colors) in translational (a, c, e) and rotational (b, d, f) motions. Ellipsoids in the same cluster have the same color. (a, b) The same frame at $\phi=0.70$ (supercooled liquid); (c, d) The same frame at $\phi=0.77$ (orientational glass); (e, f) The same frame at $\phi=0.81$ (glass) with $\sim$5500 particles.} \label{fig:cluster}
\end{center}
\end{figure}
The two glass transitions can be further confirmed from the spatial distribution of the fastest-moving particles which characterizes the structural relaxation and dynamic heterogeneity~\cite{Weeks00}. In Fig.~\ref{fig:cluster}, the fastest-moving 8\% of the particles is labeled in colors because the non-Gaussian long tail of the distribution of $\Delta x(t^*)$ covers about 8\% of the population. Here $t^*$ corresponds to the maximum of $\alpha_2$~\cite{Weeks00}. Different choices of $t$ and the percentage yield the similar results. Neighboring fastest-moving ellipsoids form clusters and are labeled using the same color. Here two ellipsoids are defined as neighbors if they overlap after being expanded by 1.5 times and their closest distance does not intersect a third particle. In the supersaturated liquid, most fast particles were strongly spatially correlated and formed large extended clusters, see Figs.~\ref{fig:cluster} and Fig.~S3 in the SM. This demonstrates the $\alpha$-relaxation occurs by cooperative particle motion in both the translational and rotational degrees of freedom: when one particle moves, another particle moves closely following the first. The colloidal glasses, in contrast, show no discernible $\alpha$-relaxation, and the fastest particles in $\beta$-relaxation are randomly dispersed without forming large clusters~\cite{Weeks00}, as observed in the 3D glass transition of colloidal spheres~\cite{Weeks00}. Figure~\ref{fig:cluster} clearly depicts three regimes: both the translational and rotational fast particles are distributed heterogeneously with large clusters at $\phi<0.72$; the rotational fast particles are dispersed homogenously while the translational fast particles form large clusters at  $0.72<\phi<0.79$; and both types of fast particles are dispersed homogenously at $\phi>0.79$.

The spatial distributions of translational and rotational fast-particle clusters were anticorrelated. Figures~\ref{fig:cluster}(a,c) show that most translational fast particles belonged to a few large ribbon-like clusters aligned with their long axes within the pseudo-nematic domains. In contrast, the clusters of rotational fast particles formed branch-like structures extending over several small domains around the domain boundaries, see Fig.~\ref{fig:cluster}(b). Fast rotational ellipsoids moved between domains by cooperative rotational motion. This demonstrates that the nematic order within a domain facilitates translational relaxation while the orientational disorder near domain boundaries promotes rotational relaxation. Fast translational particles are responsible for the out-of-cage diffusion, while fast rotational particles are responsible for domain transformations such as splitting, merging and rotating. All the phases in Figs.~\ref{fig:cluster}(a-f) contain some isolated fast translational and rotational particles; they are mainly distributed at the domain boundaries with random orientations.

\begin{figure}
\begin{center}
\includegraphics[width=1\columnwidth]{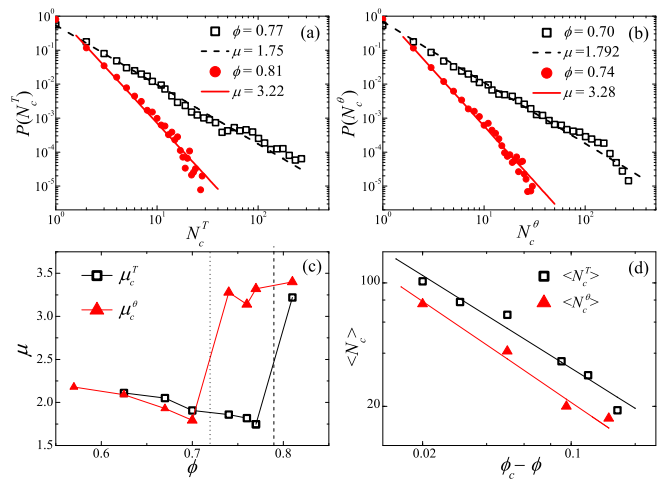}
\caption{(color online) The probability distribution functions for the cluster size of (a) translational and (b) rotational fastest-moving particles. The lines are the best fits of $P(N_c)\sim N_c^{-\mu}$. (c) The fitted exponents $\mu^{\theta}$ for rotational motions and $\mu^{_T}$ for translational motions. The vertical dotted and dashed lines represent the glass transitions for rotational and translational motions respectively. (d) The weighted mean cluster size $\langle N_c \rangle \sim (\phi_c-\phi)^{-\eta}$ where $\phi_c^{\theta}=0.71$ and $\phi_c^T=0.79$.} \label{Fig:cluster1}
\end{center}
\end{figure}
The cluster sizes of the fast particles, $N_c$, exhibit a power-law distribution $P(N_c)\sim N_c^{-\mu}$ as shown in Figs.~\ref{Fig:cluster1}(a, b). The fitted exponents $\mu$ for translational and rotational motions change dramatically near their respective glass transitions see Fig.~\ref{Fig:cluster1}(c). The $\mu^{\theta,T}=2.0\pm 0.2$ for supersaturated liquids is close to the $\mu^{_T}=2.2\pm 0.2$ estimated for hard spheres~\cite{Weeks00} and the $\mu^{_T}=1.9\pm 0.1$ for Lennard-Jones particles in 3D~\cite{Donati99}, while the $\mu^{\theta,T}=3.2\pm 0.1$ for glasses is close to the $\mu^{_T}=3.1$ estimated for hard spheres in 3D~\cite{Weeks00}. Hence $\mu \simeq 2.5$ might characterize such glass transitions in general.
Figure~\ref{Fig:cluster1}(d) shows the weighted mean cluster size $\langle N_c \rangle = \sum N_c^2 P(N_c)/\sum N_c P(N_c)$ \cite{Weeks00,Donati99} at different densities. Both $\langle N_c^{\theta} \rangle$ and $\langle N_c^{_T}\rangle$ diverge on approaching the corresponding $\phi_c$: $\langle N_c \rangle \sim (\phi_c-\phi)^{-\eta}$ with fitted $\eta^{\theta}=0.81$ and $\eta^{_T}=0.75$, indicating growing cooperative regions of mobile particles. Similar scaling and $\eta^{_T}$s have been observed in a Lennard-Jones system \cite{Donati99}, but the mechanism is not clear.

We did not observed nematic phase or semetic domains found in 3D spherocylinders \cite{Ni10} because 1) the elliptical shape facilitates particles changing orientation and forming branch-like structures at high densities \cite{Narayan06}; 2) The 5.6\% polydispersity promotes glass formation. 3) Long-wavelength fluctuations are stronger in 2D than in 3D, which can more easily break the long-range order as described by Mermin-Wagner theorem. Ellipsoids with $p\sim 6$ appeared to be good glass formers, which can easily preempt any isotropic-nematic (IN) phase transition~\cite{Cuesta90}. In contrast, the glass transition can be preempted by crystallization for $p\simeq 1$ in 2D, or by an IN transition for rods with $p\gtrsim 25$ in 3D \cite{Yatsenko08}.

All of the measurements consistently showed that the glass transitions for ellipsoids with $p=6$ confined between two walls are at $\phi_c^{\theta}=0.72$ for rotational motion and at $\phi_c^{_T}=0.79$ for translational motion. For longer ellipsoids with $p=9$ ($a = 5.9~\mu$m, $b=c=0.65~\mu$m), $\phi_c^{\theta}= 0.60\pm 0.02$ and $\phi_c^{_T}= 0.72\pm 0.02$ were observed in the two-wall confinement. This suggests that the intermediate regime between $\phi_c^{\theta}$ and $\phi_c^{_T}$ increases with the aspect ratio, which could be the reason why such an intermediate regime has not been observed in previous 3D simulations of ellipsoids with small aspect ratios~\cite{Pfleiderer08,Chong05}. We also observed the two-step glass transitions in monolayers of heavy ellipsoid sediment near one wall, but the transitions increased by 3\% area fraction because of the stronger out-of-plane fluctuations.

We conclude that colloidal ellipsoids in a 2D system exhibit two glass transitions with an intermediate orientational glass. This behavior has been predicted in 3D by MMCT but not studied in 2D before. The two glass transitions in the rotational and translational degrees of freedom correspond to inter-domain freezing and inner-domain freezing respectively. The orientational glass regime appears to increase with the aspect ratio. Approaching the glass transitions, the structural relaxation time and the mean cluster size for cooperative motion diverge $-$ typical features of a glass transition \cite{Gotze91,Gotze92,Donati99}. Interestingly, the translational and orientational cooperative motions are anticorrelated in space, which has not been predicted in theory or simulation. A similar two-step glass transition has been observed in a 3D liquid-crystal system and explained as the freezing of the orientations of the pseudo-nematic domains and the freezing of the translational motion within domains~\cite{Cang03}. Here we directly observed the conjectured pseudo-nematic domains in ref.~\cite{Cang03}. These results at single-particle resolution shed new light on the formation of molecular glasses, especially at low dimensionality.

We thank Ning Xu and Penger Tong for the helpful discussion. This work was supported by the HKUST grant RPC07/08.SC04 and by GRF grant 601208.

\end{document}